# Structural, magnetic, and electrical properties of collinear antiferromagnetic heteroepitaxy cubic Mn$_3$Ga thin films

Hyun-Woo Bang[1], Woosuk Yoo[1], Chungman Kim[1], Sunghun Lee[2], Jiyeong Gu[3], Yunchang Park[4], Kyujoon Lee[5] and Myung-Hwa Jung[1,*]

[1]*Department of Physics, Sogang University, Seoul 04107, Republic of Korea*
[2]*Department of Physics and Astronomy, Sejong University, Seoul 05006, Republic of Korea*
[3]*Department of Physics and Astronomy, California State University LongBeach, Long Beach, CA 90840, USA*
[4]*National nanofab center, Daejeon 34141, Republic of Korea*
[5]*Institute of Physics, Johannes Gutenberg University Mainz, Mainz 55128, Germany*

Abstract

Although a cubic phase of Mn$_3$Ga with an antiferromagnetic order has been theoretically predicted, it has not been experimentally verified in a bulk or film form. Here, we report the structural, magnetic, and electrical properties of antiferromagnetic cubic Mn$_3$Ga (C-Mn$_3$Ga) thin films, in comparison with ferrimagnetic tetragonal Mn$_3$Ga (T-Mn$_3$Ga). The structural analyses reveal that C-Mn$_3$Ga is hetero-epitaxially grown on MgO substrate with the Cu$_3$Au-type cubic structure, which transforms to T-Mn$_3$Ga as the RF sputtering power increases. The magnetic and magnetotransport data show the antiferromagnetic transition at $T_N$ = 400 K for C-Mn$_3$Ga and the ferrimagnetic transition at $T_C$ = 820 K for T-Mn$_3$Ga. Furthermore, we find that the antiferromagnetic C-Mn$_3$Ga exhibits a higher electrical resistivity than the ferrimagnetic T-Mn$_3$Ga, which can be understood by spin-dependent scattering mechanism.

Keywords: Heusler compound, antiferromagnetic cubic material, magnetic transition, hetero-epitaxy

Introduction

Most of Heusler compounds crystallize in the cubic L2$_1$ structure, where special attention has been focused on half-metallic ferromagnetism to exhibit a metallic behavior in one spin channel and an insulating behavior in the other spin channel, resulting in complete spin polarization of electrons at the Fermi level [1-3]. Among them, Mn-based Heusler compounds have been received of great interest due to the tetragonally distorted structure showing half-metallic ferrimagnetism [4-13], which has an advantage of low saturation magnetization requisite for spin-transfer-torque based spin devices [14-16]. The magnetization of tetragonal Mn$_3$Ga vanishes over a wide range of temperature because the magnetic moments of two Mn sublattices are antiferromagnetically aligned with different magnitude [4,5,13]. However, the tetragonal distortion tends to destroy the compensated ferrimagnetism as well as the half-metallic behavior [2,13]. In theory, a cubic phase of Mn$_3$Ga is predicted to display half-metallicity with collinear antiferromagnetic order, anticipating both complete spin polarization and zero net magnetic moment [11-13,17-22], which are meaningful to lower energy loss in spintronic device applications. However, the cubic phase of Mn$_3$Ga has not been experimentally verified yet in a bulk or film form. There has been only one report on nanostructured ribbons of cubic Mn$_3$Ga phase, built with nano-sized particles, which are made by quenching method of arc melting and melt spinning [7]. Unfortunately, the cubic antiferromagnetic phase is not thermally stable and undergoes phase transitions to tetragonal ferrimagnetic phase at 600 K and to hexagonal antiferromagnetic phase at 800 K.

In the present work, we have successfully fabricated hetero-epitaxial Mn$_3$Ga films with stable cubic phase by using RF magnetron sputtering method. We report the structural, magnetic, and electrical properties of the cubic Mn$_3$Ga (C-Mn$_3$Ga), in comparison with the tetragonal phase (T-Mn$_3$Ga). The structural analyses reveal that C-Mn$_3$Ga deposited with low RF power crystallizes in the disordered Cu$_3$Au-type structure, and it transforms to T-Mn$_3$Ga as the power increases. We find the antiferromagnetic transition at $T_N$ = 400 K for C-Mn$_3$Ga, while the ferrimagnetic transition at $T_C$ = 800 K for T-Mn$_3$Ga. Furthermore, the electrical resistivity is higher in the antiferromagnetic phase of C-Mn$_3$Ga than that in the ferrimagnetic phase of T-Mn$_3$Ga.

Experimental Details

The films of Mn$_3$Ga were deposited on MgO(001) substrate using RF magnetron sputtering with a base pressure of $1.0 \times 10^{-6}$ Torr. The RF power was varied from 10 W to 55 W with a constant substrate temperature of 400°C and Argon pressure of 2mTorr during the deposition. The crystal structure of the samples was determined by using X-ray diffraction (XRD Bruker AXS D8 Discover diffractometer using Cu $K\alpha$ radiation). In addition, high-resolution transmission electron microscopy (HR-TEM FEI Tecnai G$^2$ F30 S-TWIN) and transmission electron diffraction (TED) were used for detailed structural investigation of Mn$_3$Ga with MgO substrate. The surface morphology and relative Mn composition were measured using the scanning electron microscope (SEM) and electron dispersive x-ray spectroscopy (EDX JEOL JSM-6700F). The magnetic properties and electron-transport properties were measured using a superconducting quantum interference device-vibrating sample magnetometer (SQUID-VSM Quantum Dsign MPMS) where the magnetic field was swept from -7 T to 7 T and temperature range 2 ~ 300 K. The temperature dependence of magnetization with high temperature was measured from 300 K to 800 K by using a physical property measurement system (VSM PPMS).

Results and Discussion

The structural evolution with varying the deposition conditions has been investigated by the X-ray diffraction (XRD) measurements. Figure 1(a) shows the XRD patterns of Mn$_3$Ga films grown with various deposition powers of the RF magnetron sputtering system. As the RF power decreases from 50 to 25 W, the D0$_{22}$ tetragonal phase of Mn$_3$Ga (T-Mn$_3$Ga in Fig. 1(b)) is slowly transformed to the cubic Mn$_3$Ga phase (C-Mn$_3$Ga in Fig. 1(c)) with the disordered Cu$_3$Au-type (L1$_2$) structure. Besides the peaks from MgO substrate, the XRD patterns mainly show three different peaks from the samples, which are two (002) and (004) tetragonal peaks and one (002) cubic peak. For the films deposited with high powers ($P > 43$ W), there are two dominant peaks at 24.99° and 51.28° which coincide with the (002) and (004) peaks of the D0$_{22}$ tetragonal structure, respectively. For the films grown with low powers ($P < 41$ W), we observe a peak at 48.27° which is matched with the (002) peak of the disordered L1$_2$ cubic structure [7,23,24]. In the intermediate RF powers (41 W $\leq P \leq$ 43 W), a mixture of both tetragonal and cubic phases is found even though the diffraction peaks become broader than those of single phase of T-Mn$_3$Ga or C-Mn$_3$Ga. It is clear that the structural phase change from the tetragonal structure to the cubic structure occurs as the RF power decreases. The lattice parameters obtained from the XRD analyses for T-Mn$_3$Ga are $c = 7.11$ Å and $a = 3.89$ Å by setting $c/a = 1.83$, which are the same values reported by other literatures [4-6,12]. For C-Mn$_3$Ga, we estimate the lattice parameter of $a = c = 3.76$ Å, which is consistent with that in the nanostructured ribbons of Mn$_3$Ga proposed to have the Cu$_3$Au-type cubic structure [7]. Here, it should be pointed out that the lattice mismatch of C-Mn$_3$Ga with the MgO substrate *($a = c = 4.21$ Å)* is larger than that of T-Mn$_3$Ga, and the C-Mn$_3$Ga phase is not a stable phase in nature. Nevertheless, we obtain epitaxial films of both T-Mn$_3$Ga and C-Mn$_3$Ga, which are demonstrated by the in-plane phi scans. The representative plots are shown in Figs. 1(d) and (e), where the peaks are observed at 90 degree intervals indicating four-fold symmetry. The results suggest that both T- and C- Mn$_3$Ga films on MgO substrate are epitaxially grown with high quality.

To elucidate the epitaxy of C-Mn$_3$Ga, we have performed the transmission electron microscopy (TEM) and transmission electron diffraction (TED) measurements. Fig. 2(a) shows the TEM image of C-Mn$_3$Ga deposited with the RF power of 25 W. The two different layers of MgO(100) substrate and C-Mn$_3$Ga sample are clearly distinguished in the TEM image. The thickness of C-Mn$_3$Ga is about 10 nm. In Fig. 2(b), we observe two distinct TED patterns corresponding to (200) orientation of C-Mn$_3$Ga and MgO with four-fold symmetry, in consistent with the result of XRD pi scan experiments. The lattice parameters from the TED results are estimated to be $a = c = 3.78$ Å and 4.21 Å for C-Mn$_3$Ga and MgO, respectively, which also agree well with the values from the XRD results. The lattice mismatch between C-Mn$_3$Ga and MgO is about 10.2%, which is too large to consider the epitaxial growth of C-Mn$_3$Ga phase. In order to investigate the microstructure at the interface between C-Mn$_3$Ga and the MgO, we have taken a magnified image at the interface. Figure 2(c) shows the magnified TEM image of the area in the red box of Fig. 2(a). The atoms of C-Mn$_3$Ga are marked with red circles in the upper part and the atoms of MgO are marked with yellow circles in the lower part. It is clearly seen that there is an atomic stacking ratio of 10:9 between C-Mn$_3$Ga and MgO at the interface with small dislocation of a few atomic layers. This kind of growth mechanism is well known in hetero-epitaxial thin films with a large lattice mismatch of more than 9% [25-27]. In the hetero-epitaxial growth, the films are grown by domain-matching epitaxy. In the scanning electron microscopy (SEM) images of the surface morphology shown in Figs. 2(d) and (e), the domain boundaries are observed in C-Mn$_3$Ga, compared with the flat surface in T-Mn$_3$Ga. This difference in domain structure would be a natural feature when considering the domain-matching epitaxial growth.

The most prominent change in the crystal structure is clearly seen in the magnetism. We probe the magnetic transition temperatures of two T-Mn$_3$Ga and C-Mn$_3$Ga phases by measuring high-temperature magnetization up to 820 K. Figure 3(a) shows the temperature dependence of remanent magnetization for T-Mn$_3$Ga, measured when the external magnetic field is removed after field cooling. The magnetization abruptly increases below $T_C$ = 800 K corresponding to the ferrimagnetic transition temperature of T-Mn$_3$Ga. For C-Mn$_3$Ga, on the other hand, we have measured the temperature dependent magnetization in an applied magnetic field of 1 kOe because of no remanence. In Figure 3(b), the magnetization exhibits a sharp peak at $T_N$ = 400 K, which is close to the temperature proposed as an antiferromagnetic transition temperature in the cubic phase of Mn$_3$Ga [7,8,28-30]. These magnetic data recorded in thin films are different from those taken with nano-ribbons [8], where the antiferromagnetic cubic Mn$_3$Ga undergoes multiple magnetic and structural transitions to ferrimagnetic tetragonal phase at 600 K and to antiferromagnetical hexagonal phase at 800 K, and they are thermally irreversible. The irreversibility has been explained by the unstable cubic phase of Mn$_3$Ga in nature because the cubic phase is obtained only by a nonequilibrium synthesis process such as rapid quenching from a very high temperature. However, in our case of a thin-film form, we obtain a quite stable cubic phase of Mn$_3$Ga, which may be related to a strain effect of the MgO substrate. Notably, we obtain two different stable phases of ferrimagnetic T-Mn$_3$Ga and antiferromagnetic C-Mn$_3$Ga simply by changing the RF deposition power. As aforementioned, the lattice mismatch between Mn$_3$Ga sample and MgO substrate is large (~ 10.2%). When such materials are deposited on the substrate with a large lattice mismatch, higher kinetic energy is necessary to overcome the energy barrier of metastable state and achieve the stable state. In the present case, the metastable state is cubic phase of $Mn_3Ga$ and the stable state is the tetragonal phase of $Mn_3Ga$, and the higher deposition power means higher kinetic energy giving rise to the deposition of stable T-$Mn_3Ga$ phase. On the other hand, the lower deposition power could result in the growth of metastable C-$Mn_3Ga$ phase [31-34].

Figs. 3(c)-(e) show the magnetization $M(H)$ curves at room temperature for the three typical phases of T-$Mn_3Ga$, M-$Mn_3Ga$, and C-$Mn_3Ga$. The magnetic fields are applied perpendicular and parallel to the film plane, and the background signals from the diamagnetic substrate are subtracted. In Fig. 3(c), T-$Mn_3Ga$ exhibits clear hysteresis loop in the out-of-plane configuration, indicating the perpendicular magnetic anisotropy found in a tetragonal system [4,10]. The saturation magnetization and anisotropy constant values are extracted to be $M_S = 220$ emu/cc and $K_{eff} = 0.97 \times 10^6$ J/m$^3$, which are consistent with previous results [4,10,12,13]. In Fig. 3(d) for M-$Mn_3Ga$ deposited with an intermediate power of 43 W, which is a mixture of the cubic and tetragonal phases, the saturation magnetization is approximately three times lower than that of T-$Mn_3Ga$. The low saturation magnetization is due to the appearance of the cubic phase of $Mn_3Ga$. In other words, the total volume of ferromagnetic component decreases compared to the pure T-$Mn_3Ga$ phase. In Fig. 3(e), C-$Mn_3Ga$ shows no hysteresis behavior but only a linear field dependence, demonstrating the antiferromagnetic order. Note that all the samples show abrupt change at low magnetic fields, which may come from small misalignment from the $c$ axis or small misorientation in lattice.

Figures 3(f)-(h) represent the Hall resistivity $\rho_{xy}(H)$ curves of T-$Mn_3Ga$, M-$Mn_3Ga$, and C-$Mn_3Ga$ obtained at room temperature for the field along the $c$ axis. The results are in good agreement with the $M(H)$ curves. We observe clear hysteresis loops for T-$Mn_3Ga$ and M-$Mn_3Ga$, whereas no hysteresis loop is found in C-$Mn_3Ga$. Here it is noteworthy that there is a slight shift of the hysteresis loop in M-$Mn_3Ga$, which is an indication of exchange bias effect. If the ferrimagnetic states of T-$Mn_3Ga$ coexist with the antiferromagnetic states of C-$Mn_3Ga$, the exchange bias effect can be expected. The shift of hysteresis loop is quite small because the magnetic field and temperature

required for the conventional exchange bias effect are too low enough to affect the exchange bias. From the high-field data with linear dependence, we calculate the carrier density of $n = 1.0 \times 10^{20}$, $1.3 \times 10^{20}$, and $1.9 \times 10^{20}$ cm$^{-3}$ for T-Mn$_3$Ga, M-Mn$_3$Ga, and C-Mn$_3$Ga, respectively. These values lie in poor metallic regime, which is necessary for later discussion on the electrical transport.

We investigate the temperature dependence of electrical resistivity $\rho(T)$ for C-, M-, and T-Mn$_3$Ga. The results are displayed in Fig. 4(a). Since M-Mn$_3$Ga can have dominant contributions from the different volume of mixed C- and T-Mn$_3$Ga phases, we select two different M-Mn$_3$Ga films deposited with the RF powers of 43 and 41 W, which correspond to tetragonal- and cubic-phase dominant samples, respectively. As shown in Fig. 4(a), the electrical resistivity of T-Mn$_3$Ga is distinct from that of C-Mn$_3$Ga, i.e., they display very different behavior not only in temperature dependence but also in magnitude. T-Mn$_3$Ga displays metallic behavior, C-Mn$_3$Ga exhibits semiconducting behavior, and M-Mn$_3$Ga shows the intermediate behavior depending on the dominant phase; $\rho(T)$ of the tetragonal-phase dominant M-Mn$_3$Ga (43 W) is close to that of T-Mn$_3$Ga and $\rho(T)$ of the cubic-phase dominant M-Mn$_3$Ga (41 W) is close to that of C-Mn$_3$Ga. The resistivity values are also changed sequentially depending on the structural change. According to the carrier density estimated from the Hall measurements, C-Mn$_3$Ga has more carriers than T-Mn$_3$Ga, so that $\rho(T)$ of C-Mn$_3$Ga must be lower than that of T-Mn$_3$Ga. However, we observe the opposite behavior in experiment. The carrier mobility estimated from the carrier density and resistivity value is 1,400, 950, and 510 cm$^2$/Vs for T-Mn$_3$Ga, M-Mn$_3$Ga, and C-Mn$_3$Ga, respectively, suggesting that the electrical resistivity is governed mostly by the carrier mobility. One possible explanation for the difference between T-Mn$_3$Ga and C-Mn$_3$Ga is the effect of grain boundary scattering on the electron transport. As shown in the SEM images in Fig. 2(d) and (e), more grain boundaries exist in C-Mn$_3$Ga, resulting in the reduced carrier mobility and the increased electrical resistivity. However, this grain boundary effect cannot explain the intermediate behavior of M-Mn$_3$Ga. Another explanation can be the spin-dependent scattering mechanism, which is normally discussed in giant magnetoresistance effect [35-38]. The electrical resistance is larger for the collinear antiferromagnetic spin configuration.

Since the electrical transport is strongly affected by the magnetic order in magnetic materials, it is useful to compare magnetoresistance with magnetization. As displayed in Fig. 4(b) and (c), clear two peaks in the magnetoresistance of T-$Mn_3Ga$ coincides with the sharp peaks of differential magnetization data at $H_C$ = 15 kOe. On the other hand, no anomaly is found in C-$Mn_3Ga$, where the magnetoresistance changes by the order of 0.1% of the total resistance.

Conclusions

Our results show that the cubic phase of $Mn_3Ga$ can be stabilized and manipulated by reducing the deposition power in RF magnetron sputtering. Notably, depending on the crystal structure of $Mn_3Ga$, two distinct magnetic phases have been observed experimentally; cubic $Mn_3Ga$ (C-$Mn_3Ga$) and tetragonal $Mn_3Ga$ (T-$Mn_3Ga$). The XRD and TEM analyses show that C-$Mn_3Ga$ is hetero-epitaxially grown on MgO substrate in spite of large lattice mismatch. From the magnetic field and temperature dependent magnetization measurements, we confirm C-$Mn_3Ga$ to be antiferromagnetic with $T_N$ = 400 K and T-$Mn_3Ga$ to be ferrimagnetic with $T_C$ = 800 K. The electrical transport data provide poor metallicity in C-$Mn_3Ga$, which can be understood by spin-dependent scattering in collinear antiferromagnetic spin structure. These results enlarge the family of Heusler compounds and pave a new way to the engineering of new antiferromagnetic material for future spintronic device applications.

ACKNOWLEDGEMENT

This work was supported by the National Research Foundation of Korea (NRF) grant funded by the Korea government (No. 2016M3A7B4910400, 2017R1A2B3007918).

**References**

1) T. Graf, C. Felser, and S. S. P. Parkin, Prog. Solid State Chem. **39**, 1 (2011).

2) L. Wollmann, A. K. Nayak, S. S. P. Parkin, and C. Felser, Annu. Rev. Mater. Res. **47**, 247 (2017).

3) F. Casper, T. Graf, S. Chadov, B. Balke and C. Felser, Sci. Technol. **27**, 063001 (2012).

4) H. Kurt, K. Rode, M. Venkatesan, P. Stamenov, and J. M. D. Coey, Phys. Rev. B **83**, 020405(R) (2011).

5) E.Krén, and G. Kádár, Solid State Commun. **8**, 1653 (1970).

6) H. Niida, T. Hori, H. Onodera, Y. Yamaguchi, and Y. Nakagawa, J. Appl. Phys. **79**, 5946 (1996).

7) P. Kharel, Y. Huh, N. A-. Aqtash, V. R. Shah, R. F. Sabirianov, R. Skomski, and D. J. Sellmyer, J. Phys.: Condens. Matter **26**, 126001 (2014).

8) T. Hori, Y. Morii, S. Funahashi, H. Niida, M. Akimitsu, and Y.Nakagawa, , Physica B **213&214**, 354 (1995).

9) A. Bedoya-Pinto, C. Zube, J. Malindretos, A. Urban, and A. Rizzi, Phys. Rev. B **84**, 104424 (2011).

10) H.-W. Bang, W. Yoo, Y. Choi, C.-Y. You, J.-I. Hong, J. Dolinšek, and M.-H. Jung, Curr. Appl. Phys. **16**, 63 (2016).

11) S. Wurmehl, H. C. Kandpal, G. H. Fecher, and C. Felser, J. Phys.: Condens. Matter **18**, 6171 (2006).

12) J. Winterlik, B. Balke, G. H. Fecher, and C. Felser, Phys. Rev. B **77**, 054406 (2008).

13) B. Balke, G. H. Fecher, J. Winterlik, and C. Felser, Appl. Phys. Lett. **90**, 152504 (2007).

14) J. C. Slonczewski, J. Magn. Magn. Mater. **159**, L1 (1996).

15) J. C. Slonczewski, J. Magn. Magn. Mater. **247**, 324 (2002).

16) Y. M. Huai, AAPPS Bulletin **18**, 33 (2008).

17) G. Y. Gao, and K. -L. Yao, Appl. Phys. Lett. **103**, 232409 (2013).

18) J. Kübler, J. Phys.: Condens. Matter **18**, 9795 (2006).

19) S. V. Faleev, Y. Ferrante, J. Jeong, M. G. Samant, B. Jones, and S. S. P. Parkin, Phys. Rev. Materials **1**, 024402 (2017).

20) L. Wollmann, S. Chadov, J. Kübler, and C. Felser, Phys. Rev. B **92**, 064417 (2015).

21) L, Wollmann, S. Chadov, J. Kübler, and C. Felser, Phys. Rev. B **90**, 214420 (2014).

22) T. Graf, J. Winterlik, L. Müchler, G. H. Fecher, C. Felser, and S. S. P. Parkin, Handbook of Magnetic Materials **21**,51 (2013).

23) T. Graf, F. Casper, J. Winterlik, B. Balke, G. H. Fecher, and C. Felser, Z. Anorg. Allg. Chem. 635, 976 (2009).

24) H. Kurt, K. Rode, P. Stamenov, M. Venkatesan, Y.-C. Lau, E. Fonda, and J. M. D. Coey, Phys. Rev. Lett **112**, 027201 (2014).

25) S. Kaneko, H. Funakubo, T. Kadowaki, Y. Hirabayashi and K. Akiyama, Europhys. Lett. **81**, 46001 (2008).

26) J. Deng, K. Dong, P. Yang, Y. Peng, G. Ju, J. Hu, G. M. Chow, and J. Chen, J. Magn. Magn. Mater. 446, 125 (2018).

27) J. Narayan, and B. C. Larson, J. Appl. Phys. **93**, 278 (2003).

28) H.-G. Meißner and K. Schubert, Z. Metallkd. **56**, 523 (1965).

29) V. Prudnikov, V. Silonov, M. Prudnikova, and S. Rodin, J. Magn. Magn. Mater. **188**, 393 (1998).

30) M. Getzlaff, Fundamentals of Magnetism (Springer, Berlin, New York, 2007).

31) C. Suryanarayana, Non-Equilibrium Processing of Materials (Elsevier, New York, 1999), Vol. 2.

32) P. F. Carcia and E. M. McCarron, Thin Solid Films 155, 53 (1987).

33) R. F. C. Farrow, Mater. Res. Soc. Symp. Proc. 37, 275 (1985).

34) A. Chaoumead, Y.-M. Sung, and D.-J. Kwak, Adv. Condens. Matter Phys. **2012**, 1 (2012).

35) S. S. P. Parkin, Phys. Rev. Lett. **71**, 1641 (1993).

36) W.H. Butler, X.-G. Zhang, D. M C Nicholson, and J. M. MacLaren, J. Magn. Magn. Mater. **151**, 354 (1995).

37) J. F. Gregg, W. Allen, K. Ounadjela, M. Viret, M. Hehn, S. M. Thompson, and J. M. D. Coey, Phys. Rev. Lett. **77**, 1580 (1996).

38) F. G. Aliev, R. Schad, A. Volodin, K. Temst, C. Van Haesendonck, Y. Bruynseraede, I. Vavra, V. K. Dugaev and R. Villar, Europhys. Lett., **63**, 888 (2003).

**Figure captions**

Figure 1. (a) X-ray diffraction patterns of Mn$_3$Ga films, where the line colors represent the diffraction patterns for different powers of 25, 30, 35, 41, 42, 43, 40, 45, and 50 W. Crystal structures of (b) tetragonal and (c) cubic Mn$_3$Ga. The red, green, and blue spheres indicate Ga, Mn I, and Mn II atoms, respectively. In-plane phi scans of (d) tetragonal and (e) cubic Mn$_3$Ga.

Figure 2. (a) Transmission electron microscope image, (b) transmission electron diffraction patterns, and (c) the magnified image of cubic Mn$_3$Ga film and MgO substrate. The red and yellow colors represent the cubic Mn$_3$Ga film and the MgO substrate, respectively. Scanning electron microscope images of (d) tetragonal and (e) cubic Mn$_3$Ga.

Figure 3. Temperature dependence of magnetization for (a) tetragonal and (b) cubic Mn$_3$Ga. Magnetic field dependence of (c-e) magnetization and (f-h) Hall resistivity data for tetragonal, mixed, and cubic Mn$_3$Ga, respectively.

Figure 4. (a) Temperature dependence of electrical resistivity for cubic, (cubic- and tetragonal-phase dominant) mixed, and tetragonal Mn$_3$Ga. (b) Magnetoresistance data for tetragonal and cubic Mn$_3$Ga, compared with (c) the first derivative of magnetization data for tetragonal Mn$_3$Ga.

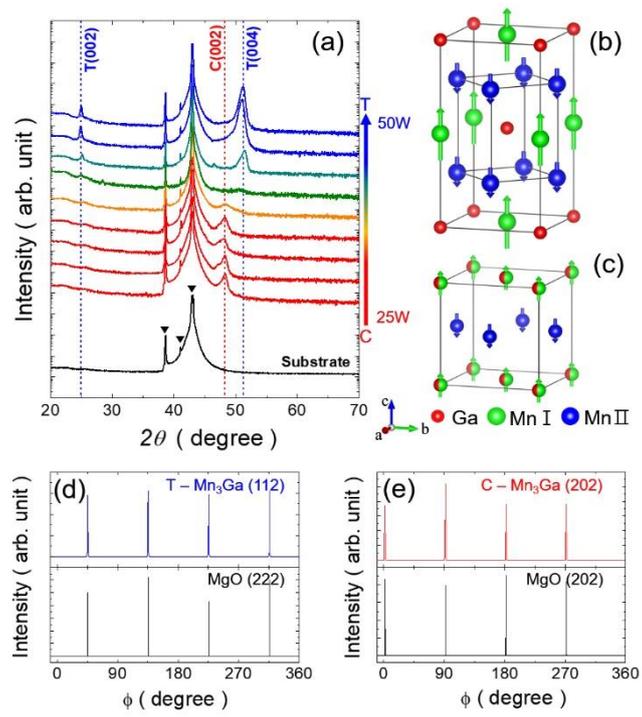

Figure 1. Bang *et al.*

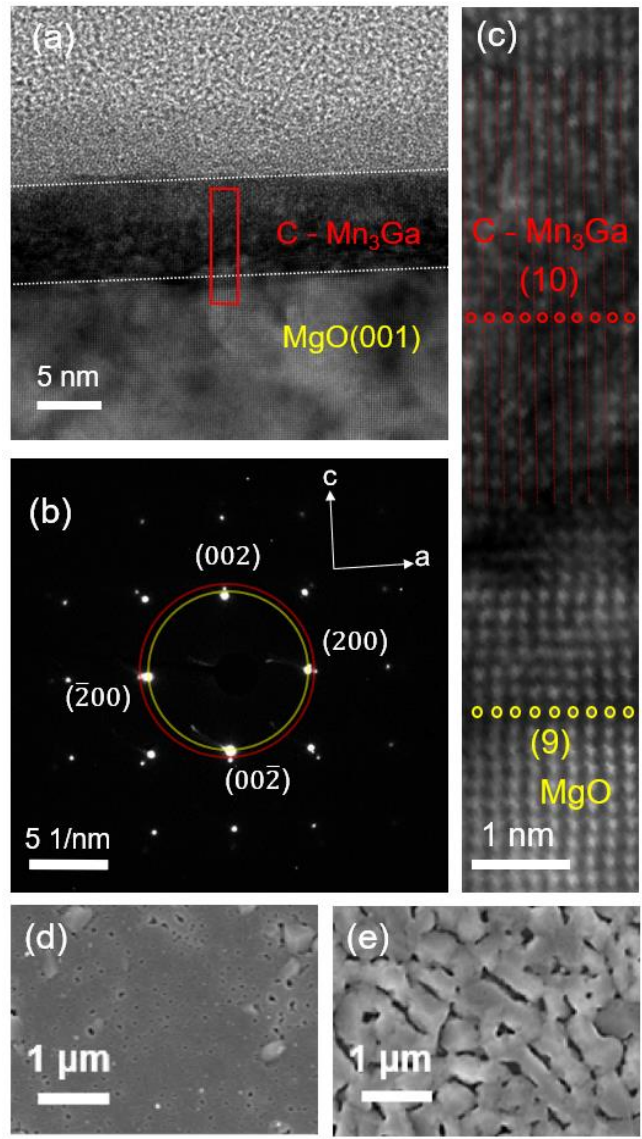

Figure 2. Bang *et al.*

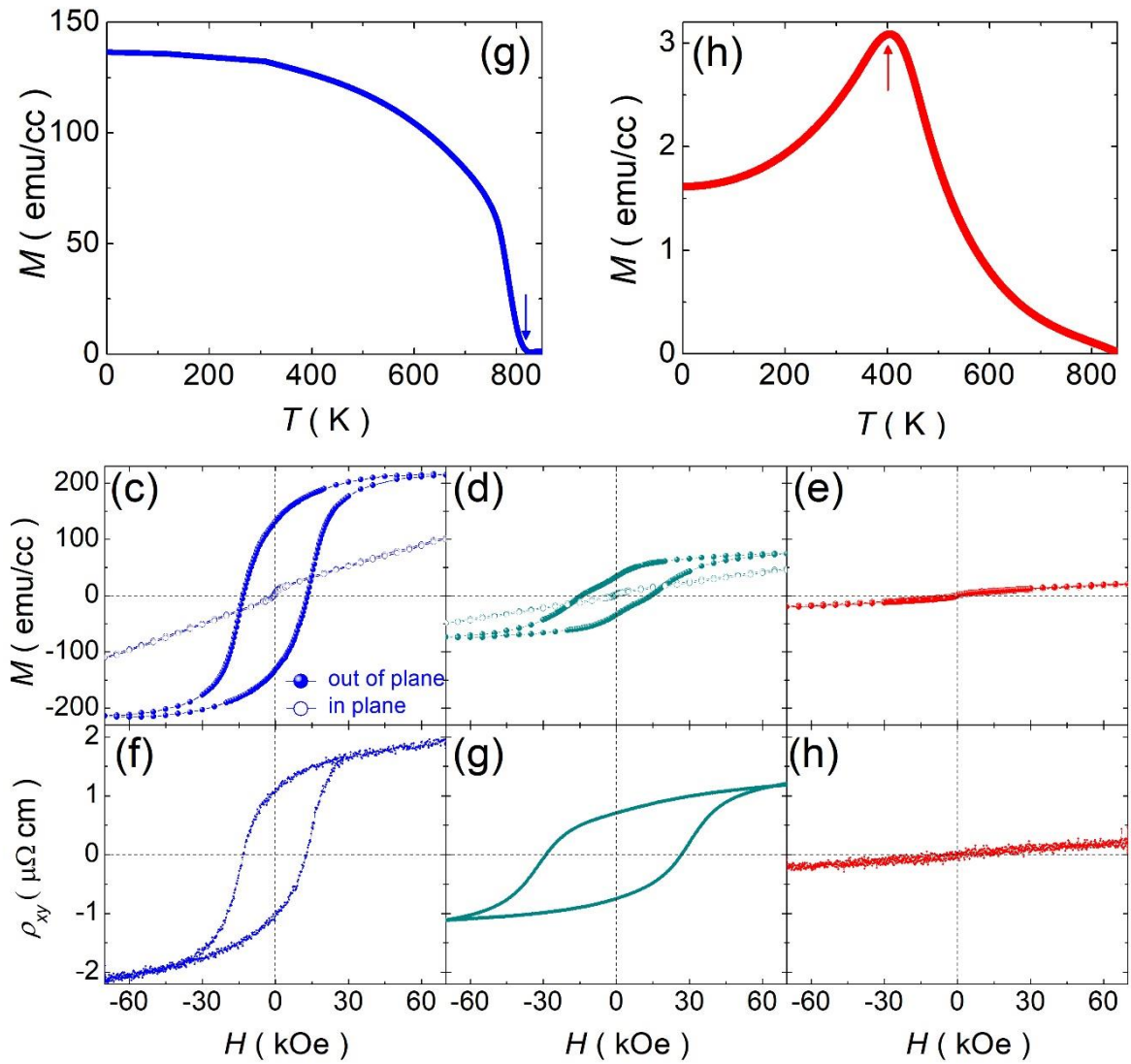

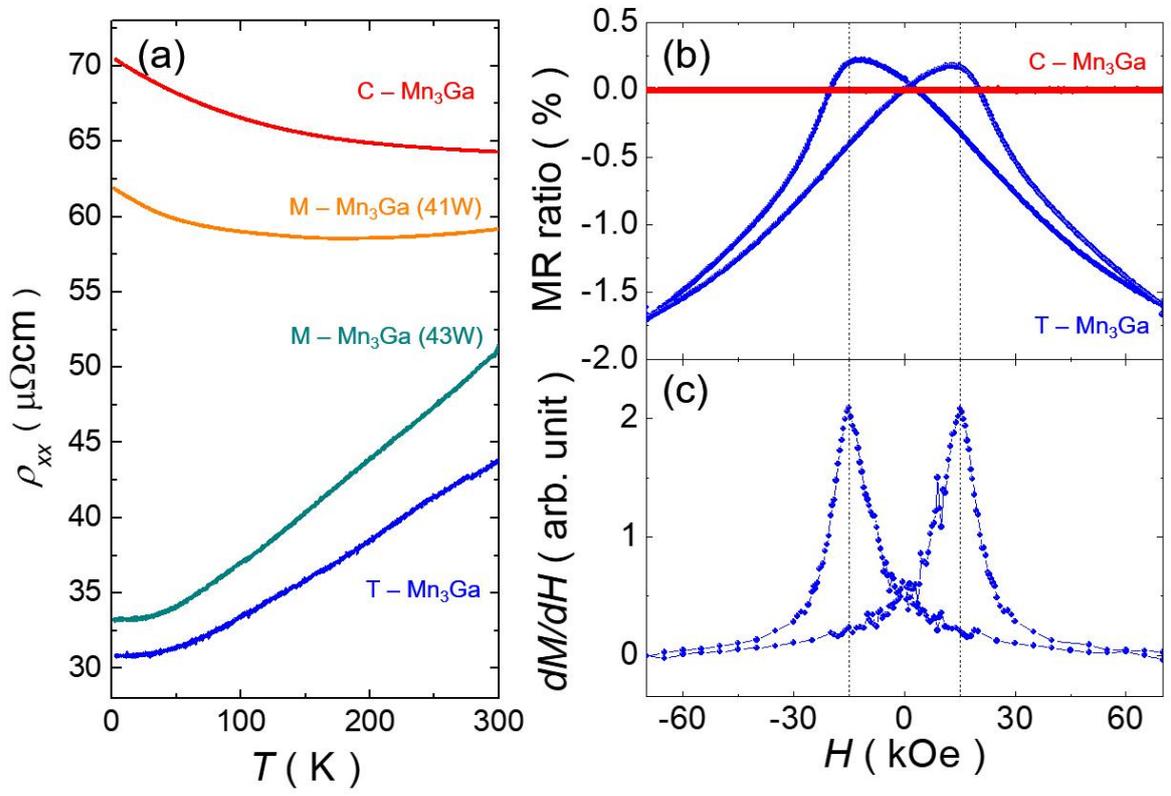

Figure 4. Bang *et al.*